\newcommand{\beq}{\begin{equation}}
\newcommand{\eeq}{\end{equation}}
\newcommand{\beqn}{\begin{eqnarray}}
\newcommand{\eeqn}{\end{eqnarray}}
\newcommand{\nn}{\nonumber}

\documentclass[12pt]{article}
\begin{document}

\title{
{\bf The stress tensor in Thermodynamics and Statistical Mechanics}}

\author{{\large{G.C. Rossi}}$^{^{a)b)*)}}$
{\large{M. Testa}}$^{^{c)d)}}$\\\\
\small $^{a)}$Dipartimento di Fisica, Universit\`a di Roma {\it Tor Vergata}\\
\small $^{b)}$INFN, Sezione di Roma 2 \\
\small Via della Ricerca Scientifica, 00133 Roma, Italy\\
\small ${^{c)}}$Dipartimento di Fisica, Universit\`a di Roma {\it La Sapienza}\\
\small $^{d)}$INFN, Sezione di Roma {\it La Sapienza}\\
\small P.le A. Moro 2, I-00185 Roma, Italy} 

\maketitle
\begin{abstract}
We prove that the stress tensor, $\tau^{ab}$, of a molecular system with arbitrary, short-range  
interactions can be point-wisely expressed as the functional derivative of the partition function with 
respect to the local deformation tensor. In this approach the set of components of $\tau^{ab}$ has a 
simple interpretation as the set of Lagrangian multipliers which need to be introduced to 
enforce the conditions relating point particle displacements to the body local deformation tensor. 
The question of the non-uniqueness of the formula for $\tau^{ab}$ is discussed. 

\end{abstract}

\vskip 2.cm
\noindent\small {$^{* )}${\it Correspondence to:} G.C. Rossi, Phone:
+39-0672594571; FAX: +39-062025259; E-mail address: rossig@roma2.infn.it}
\vfill

\newpage

\section{Introduction}
\label{sec:INTRO}

It has been shown in~\cite{I} that the invariance of the partition function of a molecular system 
under the change of variables induced by the (measure preserving) canonical 
transformation~\footnote{A sum over repeated spatial indices ($a,b,\ldots=1,\ldots,D$) is understood. 
We also frequently use the shorthand notation $\partial f(\vec r\,)/\partial r^a =\nabla^a f(\vec r\,)$.}
\beqn
&& [\{\vec r\,\},\{\vec p\,\}]\rightarrow [\{\vec r\,'\},\{\vec p\,'\}]\, ,\label{CHV}\\
&& r_i'^{a}= r_i^a+\epsilon^a\,(\vec r_i)\, ,\qquad i=1,2,\ldots, N\, ,\label{QDIFF}\\
&& p_i^{a}=\frac{\partial r_i'^{a}(\vec r\,)} {\partial r^b}\Big{|}_{\vec r=\vec r_i}\,p_i'^{b}=
\Big{[}\delta^{ab}+\frac{\partial\epsilon^a(\vec r\,)}
{\partial r^b}\Big{|}_{\vec r=\vec r_i}\Big{]}p'^{b}\, ,\label{PDIFF}\eeqn
where $\vec \epsilon\,(\vec r)$ represents the infinitesimal displacement of the body 
elements~\footnote{Boundary conditions matching those imposed to the body 
must be obeyed by the function $\vec \epsilon\,(\vec r)$.} at the point $\vec r$, implies 
for a system with short-range interactions the local equilibrium condition~\cite{LL1} 
\beq
\nabla^b \tau^{ab}(\vec r)+ {\cal{F}}^a_{\rm{ext}} (\vec r)=0 
\label{EQC}\eeq
with $\tau^{ab}(\vec r\,)$ the stress tensor at the point $\vec r$ and ${\cal{F}}^a_{\rm ext} (\vec r\,)$ 
an external force (density). The local expression of $\tau^{ab}(\vec r\,)$ in terms of the degrees of 
freedom (dof's) of the elementary constituents of the body was derived by explicitly computing 
the functional derivative of the partition function with respect to the particle displacement and 
comparing with eq.~(\ref{EQC}). The generality of the approach guarantees its validity in any statistical 
{\it ensemble}, for whichever type of (short-range) potential and boundary conditions, and in a classical 
as well in a quantum mechanical setting  (see~\cite{I} for details). 

The purpose of this brief note is twofold. First of all, after recalling how the equilibrium condition~(\ref{EQC}) 
can be deduced from purely thermodynamic considerations, we prove that to the set of stress tensor components
can be given an elegant interpretation as the set of Lagrange multipliers that are needed to enforce the relation 
between the displacement vector, $\vec\epsilon\,(\vec r\,)$, and the deformation tensor, $\eta^{ab}(\vec r\,)$, 
given by the formula~\cite{LL1} 
\beq
\eta^{ab}(\vec r\,)=\frac{1}{2}\big{[}\nabla^a\epsilon^b(\vec r\,)+
\nabla^b\epsilon^a(\vec r\,)\big{]}\, .
\label{DISDEF}\eeq
Secondly we rederive the local expression of the stress tensor for a molecular system (endowed 
with short-range interactions) by moving from the ``passive'' interpretation of the transformation~(\ref{QDIFF}) 
(followed in~\cite{I}) where eqs.~(\ref{QDIFF}) and~(\ref{PDIFF}) are seen as a mere change of variables, to an 
``active'' one where we imagine that $\vec\epsilon\,(\vec r\,)$ is the actual infinitesimal displacement of the 
body elementary constituents at the point $\vec r$. 

As a fall-out of the approach we present in this paper the question of the uniqueness of the stress tensor~\cite{UNI} 
can be neatly addressed with the conclusion that no ambiguity affects the formula~(\ref{TAU}) below. We will show, 
in fact, that there is no freedom to add to this expression any arbitrary divergenceless, 
symmetric rank-two tensor, as {\it a priori} geometrically allowed by the structure of eq.~(\ref{EQC}). 

\section{Mechanics and thermodynamics}
\label{sec:THERMO}

Let us start with a discussion of the physics of local body deformations. 
Assuming that the system is at mechanical equilibrium at fixed temperature,  
the principle of virtual works~\cite{LL2} ensures us that the work done 
by the body under an infinitesimal local deformation is zero. 
Furthermore, if the deformation transformation is reversible, the variation 
of the free energy will be equal to minus the work done by the body~\cite{LL3}. 
In these circumstances one can write the variation of the free energy of the system 
under an infinitesimal (reversible) deformation, as a function of the particle 
displacement vector and deformation tensor, in the form 
\beq
d {\cal A}(\eta,\epsilon)=-\delta_{rev} L(\eta,\epsilon) =
\int_V d^3r\, \tau^{ab} (\vec r\,)\eta^{ab}(\vec r\,) -
\int_V d^3r\, {\cal{F}}^a_{\rm ext} (\vec r\,) \epsilon^{a}(\vec r\,)\, ,
\label{FEW}
\eeq
where the first term in the last equality corresponds to the work done by the body deformation 
and the second to the work done by the external force (if there is one). As recalled above, 
the sum of the two contribution vanishes provided the displacement vector and the 
deformation tensor are related as in eq.~(\ref{DISDEF}).

The way to see what the condition of thermodynamic equilibrium implies for this 
constrained system, is to introduce Lagrange multipliers, $\lambda^{ab}$, to enforce 
eqs.~(\ref{DISDEF}), and define the ``unconstrained'' variation of  ${\cal A}$ 
\beqn 
&&d {\cal A}_{\rm uncon}(\eta,\epsilon;\lambda)=\int_V d^3r\,\tau^{ab} (\vec r\,)\eta^{ab}(\vec r\,) - 
\int_V d^3r\, {\cal{F}}^a_{\rm ext} (\vec r\,) \epsilon^{a}(\vec r\,)+\nn\\
&&-\int_V d^3r\, \lambda^{ab} (\vec r\,)\Big{[}\eta^{ab}(\vec r\,)-
\frac{1}{2}\big{[}\nabla^a\epsilon^b(\vec r\,)+
\nabla^b\epsilon^a(\vec r\,)\big{]}\Big{]}\, .
\label{FEC}
\eeqn
Imposing the vanishing of $d {\cal A}_{\rm uncon}(\eta,\epsilon;\lambda)$ immediately yields the relations  
\beqn
&&\tau^{ab} (\vec r\,)-\lambda^{ab}(\vec r\,) =0 \, ,\label{EQ1}\\
&&{\cal{F}}^a_{\rm ext} (\vec r\,) +\nabla^b \lambda^{ab} (\vec r\,)=0\, .\label{EQ2} 
\eeqn
Eqs.~(\ref{EQ1}) provide the announced interpretation of the set of stress tensor components as the set of
Lagrange multipliers which enforce the constraints~(\ref{DISDEF}). Eliminating $\lambda^{ab}$ 
between eqs.~(\ref{EQ1}) and~(\ref{EQ2}) gives back the body equilibrium condition~(\ref{EQC}).

\section{Statistical Mechanics}
\label{sec:STAT}

We now want to get an explicit expression of the stress tensor in terms of the elementary 
dof's of the system. To this end we need to display the functional dependence of the free 
energy on $\eta^{ab}$ and $\epsilon^{a}$. Upon comparing with the form of 
eq.~(\ref{FEW}), one can derive the desired formula for $\tau^{ab} (\vec r\,)$. 

The procedure outlined above can be straightforwardly implemented in Statistical Mechanics. 
Let us consider, in fact, a system interacting through the short-range potential 
${\cal{U}}[\{q\}] $ and let ${\cal{U}}_{\rm{ext}}[\{q\}]$ be a generic external potential. 
Working, for concreteness, in the {\it canonical ensemble} (but the argument that follows would 
similarly go through in the {\it micro-canonical ensemble}~\cite{I}), one has for the free energy 
the formulae 
\begin{eqnarray}
&& {\cal A}=-\frac{1}{\beta}\log {\cal Z}^0_c \, ,\label{FENL}\\
&&{\cal{Z}}^0_c=\int \prod^N (d^{D}p) \int_V \prod^N (d^{D}q)
\,\exp\Big{(}-\beta\, {\cal{H}}^0_{\rm ext}[\{q\},\{p\}]
\Big{)}\, , \label{ZETAC} \\\nn\\
&&{\cal{H}}^0_{\rm ext}[\{q\},\{p\}]= {\cal{H}}^0[\{q\},\{p\}] +
{\cal{U}}_{\rm{ext}}[\{q\}]\, ,\label{HEXTT}\\
&&{\cal{H}}^0[\{q\},\{p\}]=\sum_{i=1}^{N}\frac{({\vec{p}_{i}})^2}{2m}+
{\cal{U}}[\{q\}]\, . \label{HH}
\end{eqnarray} 
In eq.~(\ref{ZETAC}) the symbol $\prod^N (d^{D} p) \prod^N (d^{D} q)$ is a short-hand 
notation for the $D$-dimensional integration measure over the system phase space and $V$ is 
the volume of the box in which the system is contained. The superscript ``$^0$'' in the previous 
equations is to recall that they refer to the undeformed body in equilibrium.

To find the functional dependence of the free energy upon $\eta^{ab}$ and $\epsilon^a$, we 
have to provide the expression of the Hamiltonian of a system subjected to a local deformation 
(of the type indicated in eq.~(\ref{QDIFF})). To this end we first notice that under the infinitesimal 
displacement $\vec\epsilon(\vec q\,)$ the line element squared changes according to the formula~\cite{LL1} 
\beq
dq^adq^a\rightarrow dq^adq^a+2\eta^{ab}\,(\vec q\,)dq^a dq^b\, ,
\label{LEC}\eeq
with $\eta^{ab}(\vec q\,)$ related to $\epsilon^{a}(\vec q\,)$ as in eq.~(\ref{DISDEF}). 
Consequently also the kinetic energy of the system will get modified because of the addition 
of the extra contribution coming from the second term in eq.~(\ref{LEC}). In fact, from 
eq.~(\ref{LEC}) one formally gets for the modulus square of the velocity 
\beq
\frac{dq^a}{dt}\frac{dq^a}{dt}\rightarrow \frac{dq^a}{dt}\frac{dq^a}{dt}
+2\eta^{ab}\,(\vec q\,) \frac{dq^a}{dt}\frac{dq^b}{dt}\, .
\label{VEC}\eeq 
The Hamiltonian of the deformed system will thus read 
\beqn
\hspace{-1.5cm}&&{\cal H}_{\rm ext}[\{q\},\{p\};\eta,\epsilon]={\cal H}_{\rm ext}^0[\{q\},\{p\}]+\nn\\
\hspace{-1.5cm}&&-\sum_{i=1}^{N}\eta^{ab}(\vec q_i)\frac{p_i^a p_i^b}{m} - 
\frac{1}{2}\sum_{j \neq i =1}^{N}\eta^{ab}(\vec q_i)q_{ij}^a\,{\cal{F}}^b_{ij}[\{q\}]-
\sum_{i=1}^{N}\epsilon^a(\vec q_i) {\cal{F}}_{i,\rm ext}^a[\{q\}]\, ,\label{DEFH}\eeqn
where we have introduced the definitions 
\beqn
&&{{\cal{F}}}_{ij}^a[\{q\}]=-\frac{\partial{\cal{U}}[\{q\}]} {\partial q_{ij}}
\cdot \frac{q_{ij}^a}{q_{ij}}\, ,\qquad 
{\cal{F}}^a_{i,\rm{ext}}[\{q\}]=-\frac{\partial
{\cal{U}}_{\rm{ext}}[\{q\}]}{\partial q_i^a}\, ,\label{F}\\
&&\vec q_{ij}=\vec q_i-\vec q_j\, ,\qquad q_{ij}=
\sqrt{\vec q_{ij}^{\,\,2}}\, .\label{QIJ}\eeqn 
The first term in the second line in the r.h.s.\ of eq.~(\ref{DEFH}) comes directly from eq.~(\ref{VEC}) 
after passing from velocities to canonical momenta. The second and third term arise as a 
consequence of the particle displacement $q^a\rightarrow q_i^a+\epsilon^a\,(\vec q_i)$
(eq.~(\ref{QDIFF})) in ${\cal{U}}[\{q\}]$ and ${\cal{U}}_{\rm{ext}}[\{q\}]$, respectively. 
While the structure of the third term is obvious, the form of the second needs some explanation. 
First of all, we observe that we can always consider a translationally invariant interaction potential 
as a function of the set of the two-particle distances $\{q_{ij}\}$. Secondly, since $\cal U$ is 
assumed to be short-range on the macroscopic scale over which $\vec \epsilon$ can appreciably vary, 
we conclude that one can get non-vanishing contributions to the r.h.s.\ of eq.~(\ref{DEFH}) only from 
terms where all particle distances are very small (smaller than some typical microscopic length). In computing 
the variation of $q_{ij}$ under a particle displacement, we are then justified in writing 
\beqn
\hspace{-1.5cm}&&q_i^a-q_j^a\rightarrow q_i^a+\epsilon^a(\vec q_i)-q_j^a-\epsilon^a(\vec q_j)=
\nabla^b\epsilon^a(\vec q_i)(q_i^b-q_j^b)+\ldots \, ,\label{VDIST}\\ 
\hspace{-1.5cm}&&q_{ij}\rightarrow q_{ij}+\frac{1}{2}\big{[}\nabla^a\epsilon^b(\vec q_i)+
\nabla^b\epsilon^a(\vec q_i)\big{]}\frac{q_{ij}^aq_{ij}^b}{q_{ij}} \ldots = 
 q_{ij}+\eta^{ab}(\vec q_i)\frac{q_{ij}^aq_{ij}^b}{q_{ij}} \ldots\, ,\label{NDIST}
\eeqn
where dots represent terms of higher order in the differences $q_{ij}^a$, 
that we neglect. When eq.~(\ref{NDIST}) is introduced in ${\cal{U}}[\{q\}]$, the 
second term immediately emerges by expanding in the small quantity $\eta^{ab}$.

We stress that, as expected, in~(\ref{DEFH}) the body deformation is completely described by the tensor 
$\eta^{ab}$, while the displacement $\epsilon^a$ is directly coupled only to the external force. 

Inserting the Hamiltonian~(\ref{DEFH}) in the formulae for the partition function and free energy, 
one obtains the sought for functional dependence on $\eta^{ab}$ and $\epsilon^a$. 
To first order in $\eta^{ab}$ and $\epsilon^a$ one thus gets for the free energy variation 
\beqn
\hspace{-1.5cm}&&d{\cal A}(\eta,\epsilon)=\frac{1}{{\cal{Z}}^0_c}
\int \prod^N (d^{D}p) \int_V \prod^N (d^{D}q) 
\,e^{-\beta\, {\cal{H}}^0_{\rm ext}[\{q\},\{p\}]}\cdot\nn\\
\hspace{-1.5cm}&&\cdot\Big{[}\!-\!\sum_{i=1}^{N}\eta^{ab}(q_i)\frac{p_i^a p_i^b}{m} - 
\frac{1}{2}\sum_{j\neq i=1}^{N}\eta^{ab}(q_i)q_{ij}^a\,{\cal{F}}^b_{ij}[\{q\}]- 
\sum_{i=1}^{N}\epsilon^a(q_i) {\cal{F}}_{i,\rm ext}^a[\{q\}]\Big{]}\, .
\label{FF}
\eeqn
For an easy comparison to the equations of sect.~\ref{sec:THERMO} it is convenient to introduce 
in each term of the sum over the index $i$, the identity $\int_V d^3r\, \delta(\vec r -\vec q_i)=1$. 
Having done this, eq.~(\ref{FF}) can be cast in the form 
\beqn
\hspace{-1.0cm}&&d{\cal A}(\eta,\epsilon)=\nn\\
\hspace{-1.0cm}&&=\int_V d^3r \,\eta^{ab}(\vec r\,)
\Big{\langle}-\sum_{i=1}^{N}\delta(\vec r -\vec q_i)\frac{p_i^a p_i^b}{m}  - 
\frac{1}{2}\sum_{j\neq i=1}^{N} \delta(\vec r -\vec q_i) q_{ij}^a\,{\cal{F}}^b_{ij}[\{q\}]\Big{\rangle}+\nn\\
\hspace{-1.0cm}&& - \int_V d^3r\, \epsilon^{a}(\vec r\,)\Big{\langle} 
\sum_{i=1}^{N} \delta(\vec r -\vec q_i) {\cal{F}}_{i,\rm ext}^a[\{q\}]\Big{\rangle}\, ,\label{FFD}
\eeqn
where $\langle\ldots\rangle$ means {\it ensemble} average. Comparing with eq.~(\ref{FEW}), 
the expression of $\tau^{ab}(\vec r\,)$ in terms of the elementary dof's of the system is readily 
identified as the tensor that multiplies the deformation tensor in the formula for the work 
done by the system under a local deformation. One finds in this way 
\beq \tau^{ab}(\vec r\,)=-\,\Big{\langle}
\sum_{i=1}^{N}\, \delta(\vec r-\vec q_i)\Big{(}\frac{p_i^a p_i^b}{m}+ 
\frac{1}{2}\sum_{j\,(\neq i)=1}^{N}q_{ij}^a\,{\cal{F}}^b_{ij}[\{q\}]
\Big{)}\,\Big{\rangle}\, .\label{TAU}\eeq
This formula is in agreement with~\cite{I} and, once integrated over volume, with the 
expression that can be found in the classical papers of ref.~\cite{IK}.

In closing we observe that the approach we have developed allows to answer the old question of 
whether a stress tensor obeying eq.~(\ref{EQC}) is unique or not~\cite{UNI}. The question arises, 
as from a purely geometrical point of view, one could imagine to add to whichever expression of 
$\tau^{ab}$ an arbitrary divergenceless, symmetric rank-two tensor, still fulfilling eq.~(\ref{EQC}).
However if, as we advocate in this paper, $\tau^{ab}$ is identified as the tensor which multiplies 
the deformation tensor, $\eta^{ab}$, in the formula which expresses the work done by the system 
under an infinitesimal local deformation (eq.~(\ref{FEW})), such a freedom does not exist anymore.

\section{Conclusions}
\label{sec:CONCL}

In this note we have given a consistent derivation of the microscopic expression of the 
stress tensor of a body, that complies with the principles of Thermodynamics and takes 
properly into account the geometrical constraints existing between the particle displacement 
vector and the body deformation tensor. In agreement with what is known to happen in the 
case of a continuum system, we find that the possibility of defining a (local) stress tensor rests 
on the assumption that the interaction potential between the body elementary constituents 
(for the rest fully arbitrary) is ``short-range''. The discussion we give is totally general and holds 
in any {\it ensemble} and whichever the boundary conditions imposed to the system might be. 
Remarkably the whole procedure goes through also in a quantum-mechanical setting~\cite{I}. 

A consequence of the line of reasoning we have presented above is that apparently 
there is no room for any ambiguity in the expression of the stress tensor we derive, if 
$\tau^{ab}$ is identified with the tensor which multiplies the deformation tensor in the 
formula for the work done by the system under an infinitesimal local deformation. 

\vspace{.5cm}

{\bf Acknowledgments} - We thank E. Presutti for a useful discussion. We also wish to thank 
E. Tamrod for his interest in our work and the organizers of the SIAM2008 Conference 
(http://www.siam.org/meetings/ms08/) where this investigation was started.

\end{document}